\documentclass[twocolumn]{webofc}

\usepackage[varg]{txfonts}   
\usepackage{hyperref}
\usepackage{url}
\usepackage{epsfig,amsmath,graphicx,amssymb,tabularx}
\hypersetup{colorlinks=true,citecolor=blue,urlcolor=blue,linkcolor=blue}
\begin{document}
\title{Microscopic Investigation of
Fusion and Quasifission Dynamics}
%
%
\author{\firstname{Liang} \lastname{Li}\inst{1}\fnsep\thanks{\email{liliang183@mails.ucas.ac.cn}} \and
	    \firstname{Xiang-Xiang} \lastname{Sun}\inst{1}\fnsep\thanks{\email{ x.sun@fz-juelich.de}} \and
        \firstname{Lu} \lastname{Guo}\inst{1}\fnsep\thanks{\email{guolu@ucas.ac.cn}} 
}

\institute{School of Nuclear Science and Technology, University of Chinese Academy of Sciences, Beijing 100049, China
          }

\abstract{We introduce the application of Time-Dependent Hartree-Fock (TDHF) theory to two key aspects of heavy-ion reaction dynamics for producing superheavy elements: fusion and quasi-fission (QF). For fusion reactions $^{48}$Ca+$^{238}$U, the capture cross sections, fusion probabilities, and evaporation-residue cross sections are calculated using the inputs from TDHF simulations, and the results are found to be in reasonable agreement with available experimental data. 
For the QF process of $^{48}$Ca+$^{249}$Bk, we show the distribution of the fragments and investigate the impact of the tensor force, significantly enhancing the role of spherical shell effects.
}
\maketitle
\section{Introduction}
\label{intro}
The investigation of superheavy elements (SHEs) constitutes a prominent and active frontier in modern nuclear physics~\cite{hamilton2013,oganessian2015a,hofmann2016}. 
Up to now, SHEs up to $Z = 118$ have been produced. This was accomplished using two kinds of reactions: cold-fusion reactions, which employed $^{208}$Pb and $^{209}$Bi targets for elements up to $Z = 113$~\cite{hofmann2000,morita2004}, and hot-fusion reactions, which utilized $^{48}$Ca projectiles on actinide targets to produce elements from $Z=114$ to $Z=118$~\cite{oganessian2006,oganessian2007,oganessian2010}. Significant experimental efforts have been made to produce superheavy nuclei with $Z = 119$ and $Z = 120$~\cite{oganessian2009,kozulin2010,hofmann2016,khuyagbaatar2020}, but no successful synthesis has yet been reported.

This lack of success stems chiefly from the prevalence of quasifission, a competing mechanism that overwhelmingly dominates in heavy systems and severely inhibits compound nucleus formation. This competition becomes particularly prominent for the synthesis of the SHEs, as the enormous Coulomb repulsion makes QF the dominant reaction channel. In contrast to fusion, which leads to the formation of a fully equilibrated compound nucleus (CN), the QF process involves a different trajectory. Although the colliding nuclei surpass the capture barrier, they do not amalgamate and subsequently re-separate into two fragments~\cite{back1983}. As a nonequilibrium mechanism, QF is characterized by several distinct features: interaction times are typically shorter than in fusion-fission, substantial mass exchange occurs, strong correlations between fragment mass and angle are observed, and a ``memory'' of the entrance channel properties is partially retained~\cite{vardaci2019}.

Conceptually, the fusion-evaporation reaction can be understood as a three-step process: (i) the capture of the projectile by the target, forming a dinuclear system; 
(ii) the formation of an equilibrated CN, 
and (iii) the de-excitation of the CN against fission. 
Most theoretical approaches provide a relatively consistent description for the capture and de-excitation process~\cite{antonenko1993,swiatecki2005,zagrebaev2015}. However, the second step, the formation of the CN, remains the most significant source of theoretical uncertainty. Predictions for this stage from various models can differ by several orders of magnitude. This crucial stage governs the evolution of the dinuclear system after capture, determining whether it proceeds to form a compact, fully equilibrated CN or re-separates into two fragments via QF. Therefore, a quantitative understanding of the fusion-QF competition, which determines the CN formation probability $P_{\mathrm{CN}}$, is essential for making reliable predictions of SHE synthesis cross sections.

To achieve a quantitative understanding, various theoretical models have been developed. Macroscopic ones can be parameterized to reproduce known cross-section data, but their reliance on adjustable parameters and limited treatment of dynamics challenge their predictive power for unmeasured systems. 
In contrast, the microscopic TDHF approach offers insights into the underlying dynamics~\cite{simenel2018,guo2018,stevenson2019,godbey2019,sun2022c,guo2018b}  
and has been successfully applied to many aspects of low-energy heavy-ion collisions, including fission~\cite{simenel2014a,goddard2015,huang2024,huang2024b}, fusion~\cite{guo2012,washiyama2015,li2019,sun2022,sun2022c,sun2023}, QF~\cite{yu2017,guo2018c,li2022,lee2024b,li2024c} and multinucleon transfer reactions~\cite{dai2014,wu2019,sekizawa2020,wu2022}. 
Although TDHF is not suitable for the full fusion-evaporation process or quantum tunneling phenomena such as sub-barrier fusion, its simulations can provide the main ingredients for coupled-channels calculations~\cite{wang2006,guo2018} and diffusion processes~\cite{sun2022,sun2023b,sekizawa2019b}. 

In this contribution, we introduce the results from Ref.~\cite{sun2023b} regarding the calculation of evaporation cross sections of the hot fusion reaction $^{48}$Ca + $^{238}$U, where the TDHF simulations are used to provide the inputs for capture and fusion processes. We combine TDHF with both the coupled-channel (CC) and fusion-by-difussion (FbD) approaches to calculate the cross sections of the three-step in fusion reactions. 
In parallel, 
we also show the quasifission process of ${}^{48}\text{Ca}+{}^{249}\text{Bk}$ from Ref.~\cite{li2022} via TDHF calculations, incorporating the orientation effects of deformed reactants.
By analyzing the resulting fragment yield distributions with different Skyrme forces, we aim to extract the specific influence of the tensor force on the QF fragments.

\section{Theoretical Framework}
\label{sec-1}
In the TDHF theory,  the dynamic process is described by the evolution of the one-body density $\hat{\rho}$, which is obtained by solving the TDHF equation
\begin{equation}
i \hbar \frac{\partial}{\partial t} \hat{\rho}=[\hat{H}(\hat{\rho}), \hat{\rho}]
\end{equation}
where $\hat{H}$ represents the single-particle Hamiltonian derived from the effective interaction. 
The TDHF theory describes the collective motion semiclassically, omitting the quantum tunneling of the many-body wave function. Consequently, when calculating capture
cross sections, a common and effective approach is to use  internuclear
potentials derived from microscopic calculations such as frozen density approximation ~\cite{guo2012}, 
density constrained TDHF (DC-TDHF)~\cite{guo2018b,godbey2019c,sun2022}, or density constrained frozen
HF (DC-FHF)~\cite{simenel2017,sun2023,sun2023b}, as the input of the coupled-channels code CCFULL~\cite{hagino1999} to calculate the penetration probability.

For hot-fusion reactions, the actinide target nuclei are often strongly deformed; therefore, the orientation effects must be explicitly considered. Refs.~\cite{sun2023b} utilize the DC-FHF method to calculate the internuclear potential directly as a function of the target's orientation.
In the DC-FHF method, the HF calculations are performed with the constraint that the total proton $p$ and neutron $n$ densities are the same as those of the ground state of the projectile and target
\begin{align}
	& \delta\left\langle H-\int d^3 r \sum_{q=p, n} \lambda_q(\mathbf{r})\left[\rho_{q}^{P}\left(\boldsymbol{r} ; \theta_P\right)+\rho_{q}^{T}\left(\boldsymbol{r}-\boldsymbol{R} ; \theta_T\right)\right]\right\rangle \notag \\
	& \quad=0
\end{align}
where $\rho^P$ and $\rho^T$ are the densities of the projectile and target for a given orientation. 
These densities are achieved by performing Eulerian rotations of Slater determinants in a three-dimensional Cartesian system~\cite{pigg2014}. $\theta_P\ \left(\theta_T\right)$ denotes the angle between the symmetry axis of the deformed projectile (target) and the collision axis. $\boldsymbol{R}$ is the vector between the centers of mass of the projectile and target. This variation procedure results in a unique Slater determinant $\Phi(\boldsymbol{R})$. The internuclear potential is then given by:
\begin{equation}
	V\left(\boldsymbol{R}; \theta_P, \theta_T\right)=\langle\Phi(\boldsymbol{R})| H|\Phi(\boldsymbol{R})\rangle\left(\theta_P, \theta_T\right)-E_P-E_T,
\end{equation}
where $E_P$ and $E_T$ are the binding energies of the projectile and target, respectively. 

The penetration probabilities $T_J\left(E_{\text {c.m.}}, \theta_T, \theta_P\right)$ for each partial wave $J$ are obtained by solving the Schrödinger equation using the incoming wave boundary condition method~\cite{hagino1999}.
The orientation-average cross-section is given as
\begin{align}
	\sigma_{\text {cap }}\left(E_{\text {c.m. }}\right)= & \int_0^1 d \cos \left(\theta_P\right) \int_0^1 d \cos \left(\theta_T\right) \notag\\
	& \times \frac{\pi}{k^2} \sum_J(2 J+1) T_J\left(E_{\text {c.m. }}, \theta_T, \theta_P\right) .
\end{align}
where $k=\sqrt{2 \mu E_{\text {c.m. }} / \hbar^2}$ is the wave number.

The fusion probability $P_{\mathrm{CN}}\left(\theta_P, \theta_T, E_{\text {c.m. }}, J\right)$ is calculated using the FbD model. The key input for this model, the injection point, is estimated from TDHF simulations. The fusion cross section $\sigma_{\text {fus}}$ is then obtained by folding $P_{\mathrm{CN}}$
\begin{align}
	\sigma_{\text {fus }}\left(E_{\text {c.m. }}\right)= & \int_0^1 d \cos \left(\theta_P\right) \int_0^1 d \cos \left(\theta_T\right) \notag \\
	& \times \frac{\pi}{k^2} \sum_J(2 J+1) T_J\left(E_{\text {c.m. }}, \theta_T, \theta_P\right) \notag \\
	& \times P_{\mathrm{CN}}\left(\theta_P, \theta_T, E_{\text {c.m. }}, J\right)
\end{align}
To facilitate comparison with experimental data, the effective fusion probability $P_{\text {fus }}$ is defined as the ratio
\begin{equation}
P_{\text {fus }}\left(E_{\text {c.m. }}\right)=\frac{\sigma_{\text {fus }}\left(E_{\text {c.m. }}\right)}{\sigma_{\text {cap }}\left(E_{\text {c.m. }}\right)}
\end{equation}

The survival probability $W_{\mathrm{sur}}\left(E_{\mathrm{CN}}^*, x, J\right)$ of the CN is calculated using a statistical model~\cite{bohr1939,feng2006,sun2023b} that considers the competition between $x$-neutron emission and fission.
The final evaporation residue (ER) cross section $\sigma_{\mathrm{ER}}$ for the $x n$ channel is
\begin{align}
	\sigma_{\mathrm{ER}}\left(E_{\mathrm{c} . \mathrm{m} .}, x\right)= & \int_0^1 d \cos \left(\theta_P\right) \int_0^1 d \cos \left(\theta_T\right) \notag \\
	& \times \frac{\pi}{k^2} \sum_J(2 J+1) T_J\left(E_{\mathrm{c} . \mathrm{m} .}, \theta_T, \theta_P\right)  \\
	& \times P_{\mathrm{CN}}\left(\theta_P, \theta_T, E_{\mathrm{c} . \mathrm{m} .}, J\right) W_{\mathrm{sur}}\left(E_{\mathrm{CN}}^*, x, J\right) \notag
\end{align}

For the QF cross section or yield for a
specific reaction channel, different impact parameters and orientations result in distinct yield contributions. These contributions are quantified using
\begin{equation}
\sigma_\lambda \propto \int_{b_{\min }}^{b_{\max }} b \mathrm{~d} b \int_0^{\frac{\pi}{2}} \mathrm{~d} \beta \sin (\theta_T) P_b^{(\lambda)}(\theta_T),
\end{equation}
where $\lambda$ denotes a specific reaction channel, and $P_b^{(\lambda)}(\theta_T)$ is the probability corresponding to a given impact parameter $b$ and orientation angle $\theta_T$. This probability is either 0 or 1 for the reaction channel $\lambda$~\cite{godbey2019,li2022,li2024c}.

\section{Results and Discussions}
\label{sec-3}
\begin{figure*}[ht]
	\centering
	\includegraphics[width=1.8\columnwidth,clip]{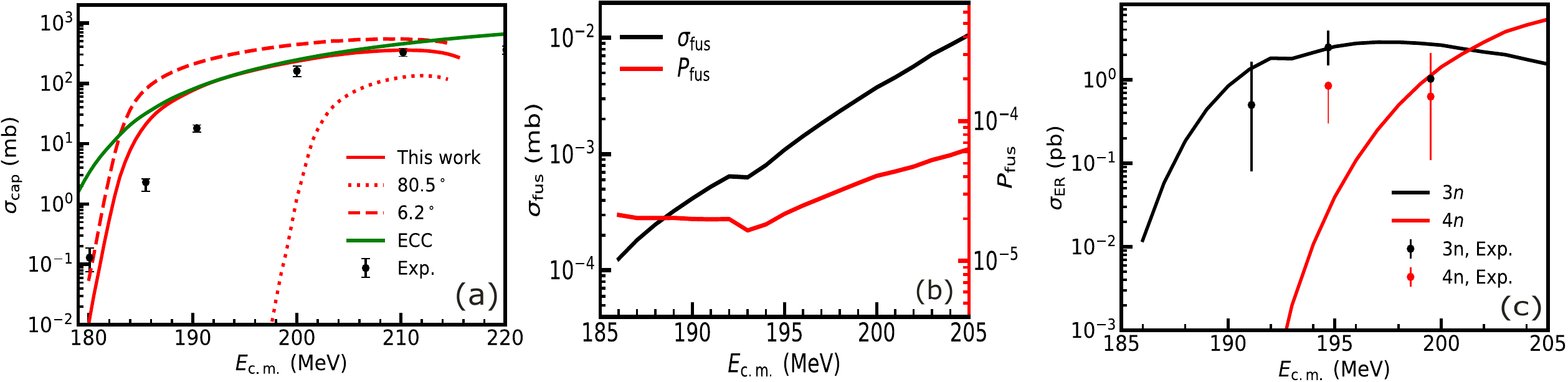}
	\caption{Theoretical results for the $^{48}$Ca + $^{238}$U reaction as a function of the center-of-mass incident energy $E_{\text{c.m.}}$. (a) Capture cross sections $\sigma_{\text{cap}}$. The orientation-averaged results (red solid line), 
tip ($\theta=9.5^{\circ}$) and side ($\theta=83.8^{\circ}$) orientations, the ECC model (green line), and experimental data (solid circles) are shown. 
(b) Orientation-averaged fusion cross sections $\sigma_{\text{fus}}$  (black line) and effective fusion probabilities $P_{\text{fus}}$ (red line). (c) Evaporation-residue cross sections $\sigma_{\text{ER}}$ for the $3n$ (black line) and $4n$ (red line) channels, compared with experimental data (points). Taken from Ref.~\cite{sun2023b}.}
	\label{fig-2}       
\end{figure*}

We first investigate the capture dynamics of the $^{48}$Ca+$^{238}$U system, in which the $^{238}$U target is prolate-deformed. Using the DC-FHF method, we obtain the internuclear potentials for seven orientations of $^{238}$U in the range $[0, \pi/2]$. 
The capture cross section was then computed from these potentials using the orientation-averaged formalism. 
As illustrated in Fig.~\ref{fig-2}(a), orientation-averaged capture cross sections $\sigma_{\text{cap}}$ (solid red line) successfully reproduce the experimental data (solid circles)~\cite{nishio2012}. 
Notably, the results demonstrate a significant improvement over the predictions of the empirical coupled-channels (ECC) model~\cite{wang2017}, particularly at sub-barrier energies below $E_{\text{c.m.}} \approx 190$ MeV. 
The plot also clearly highlights the strong dependence on the orientation of the deformed ${}^{238}$U target. The cross section for a near-tip collision ($\theta = 9.3^{\circ}$, dashed line) is substantially enhanced compared to that of a near-side collision ($\theta = 83.8^{\circ}$, dotted line). 
This enhancement is a direct consequence of the significantly lower capture barrier associated with the tip orientation.

Having established the capture cross section $\sigma_{\text{cap}}$, we next compute the orientation-averaged fusion cross section $\sigma_{\text{fus}}$ (black line in Fig.~\ref{fig-2}(b)). This calculation was performed using the FbD model \cite{swiatecki2005}. 
A key methodological strength is that the sole input parameter required by the FbD model—the injection distance—was directly extracted from our microscopic TDHF simulations.
From this computed $\sigma_{\text{fus}}$ and the previously calculated $\sigma_{\text{cap}}$, we then derived the effective fusion probability, $P_{\text{fus}} = \sigma_{\text{fus}} / \sigma_{\text{cap}}$, which is plotted as the red line. The resulting $P_{\text{fus}}$ values are found to be in the range of $(2-6) \times 10^{-4}$.  Concurrently, the calculated $\sigma_{\text{fus}}$ increases slowly with incident energy.

In the third and final stage of our calculation, we determine the evaporation-residue  cross sections, which are presented in Fig.~\ref{fig-2}(c) for the $3n$ (black line) and $4n$ (red line) emission channels. These final cross sections are obtained by interfacing our calculated fusion cross section $\sigma_{\text{fus}}$ with the survival probability $W_{sur}$. This $W_{sur}$ term is computed using a statistical model that simulates the competition between fission and neutron emission during the de-excitation of the compound nucleus. The key inputs for this stage, namely the fission barriers and neutron-separation energies, were adopted from recent microscopic-macroscopic model calculations~\cite{jachimowicz2021}. Our calculations demonstrate good agreement with the available experimental data for both channels. This overall consistency validates our comprehensive hybrid framework, which combines microscopic TDHF theory, the CCFULL calculations, the FbD model, and a statistical model as a robust and effective method for describing the complete hot-fusion reaction process.

\begin{figure*}[ht!]
	\centering
	\includegraphics[width=1.6\columnwidth]{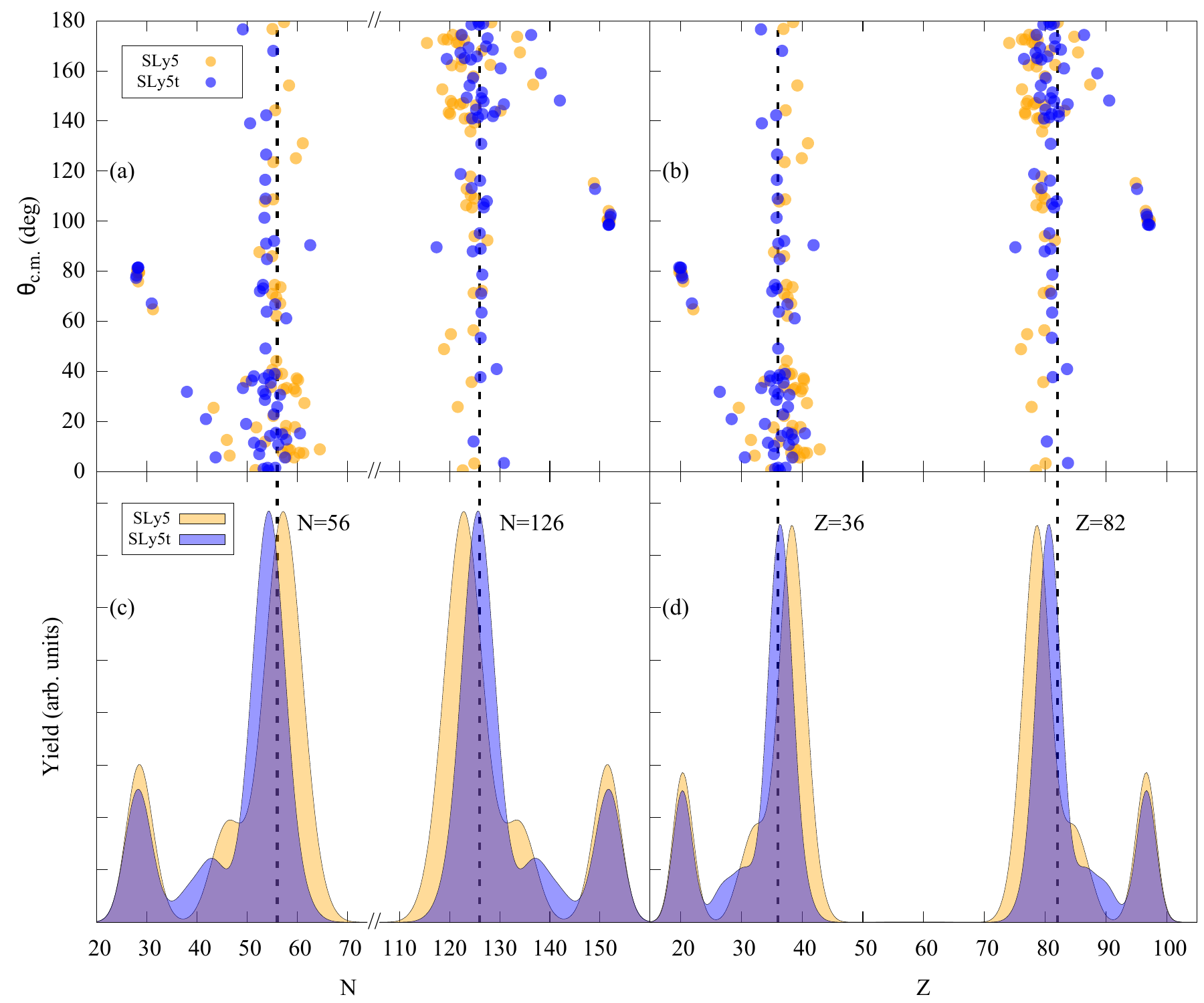}
	\caption{Quasifission reaction for the ${}^{48}$Ca+$^{249}$Bk at $E_{c.m.}=234$ MeV. (a) Scattering angle $\theta_{\text{c.m.}}$ versus fragment neutron number ($N$). (b) $\theta_{\text{c.m.}}$ versus fragment proton number ($Z$). (c) Fragment neutron number yield. (d) Fragment proton number yield. In all panels, calculations including the tensor force (SLy5t, blue points/shade) are compared with those omitting it (SLy5, orange points/shade). Vertical dashed lines mark the positions of key shell closures at $N=56, 126$ and $Z=36, 82$. Taken  from Ref.~\cite{li2022}.}
	\label{fig-3}
\end{figure*}

In the study of QF dynamics, the fragment mass-angle distributions (MADs) are crucial experimental observables. These distributions provide profound insights into the reaction mechanism, particularly by revealing the role of
quantum shell effects in shaping the characteristics of the
fragments~\cite{wakhle2014}. 
To theoretically derive the MADs and the corresponding fragment yield distributions via TDHF simulations, extensive TDHF calculations are imperative \cite{godbey2019,li2022,li2024c}. 
For the ${}^{48}\text{Ca}+{}^{249}\text{Bk}$ system, as reported in Ref.~\cite{li2022}, we systematically sampled five distinct initial orientations of the deformed ${}^{249}\text{Bk}$ target. 
For each orientation, the TDHF calculations are performed by considering a wide range of impact parameters $b$, which begins from central collisions ($b=0$) up to grazing trajectories where elastic scattering occurs.
Figure~\ref{fig-3} provides a direct comparison between calculations with the Skyrme interactions SLy5t~\cite{colo2007}, 
which includes the tensor force, 
and the SLy5 set \cite{chabanat1998a}.
The incident energy is $E_{\text{c.m.}}=234$ MeV in the center-of-mass frame. 
A clear enhancement of the neutron shell effects due to the tensor force is immediately apparent. 
The mass-angle correlation plot [Fig.~\ref{fig-3}(a)] reveals a pronounced clustering of QF fragments from the SLy5t simulation along the $N=126$ magic shell closure. 
This feature is much less distinct in the SLy5 calculation. 
This observation is quantified in the fragment yield distribution [Fig.~\ref{fig-3}(c)], where the SLy5t simulation exhibits a sharp peak centered precisely at $N=126$. In contrast, the SLy5 calculation peaks at a lower value of $N \approx 122$, indicating that the $N=126$ shell closure plays a less dominant role in the dynamics without the tensor force.

A parallel and equally significant effect is also observed for the proton. 
The SLy5t fragments [Fig.~\ref{fig-3}(b), blue points] are systematically concentrated closer to the $Z=82$ magic number than their SLy5 counterparts. 
This is corroborated by the proton yield distribution in Fig.~\ref{fig-3}(d). The SLy5t yield peaks are near $Z=82$, whereas the SLy5 calculation is centered at a lower $Z \approx 79$. Collectively, these results provide strong evidence that the inclusion of the tensor force significantly amplifies the influence of the $N=126$ and $Z=82$ shell closures, strongly driving the QF dynamics toward the production of fragments in the vicinity of the doubly-magic ${}^{208}\text{Pb}$.

\section{Summary}
\label{sec-4}
We have employed the microscopic TDHF theory, in conjunction with the coupled-channel and FbD models, to investigate the capture and fusion processes in the hot-fusion reaction $^{48}$Ca+$^{238}$U. 
For this reaction, our calculated capture cross sections are in good agreement with the experimental data. Furthermore, by accounting for the survival probability of the compound nucleus using a statistical model, our calculations successfully reproduce the experimental evaporation-residue cross sections. 
We also explored the QF dynamics in the ${}^{48}\text{Ca}+{}^{249}\text{Bk}$ reaction using TDHF simulations with and without the tensor force. 
It is revealed that the tensor force enhances the spherical shell effects. 
This enhancement is clearly manifested with more fragments produced near the magic $N = 126$ neutron shell and approaching $Z = 82$ proton shell. 

\section*{Acknowledgement}
	This work has been supported by the
	National Natural Science Foundation of China (Grants No.12375127, No.12435008, and No.12205308), the Strategic Priority Research Program of the Chinese Academy of Sciences (Grant No.XDB34010000), the Fundamental Research
	Funds for the Central Universities (Grant No. E3E46302).

\bibliography{EPJWoC.bib} 

@Article{simenel2014a,
  author    = {C. Simenel and A. S. Umar},
  title     = {{F}ormation and dynamics of fission fragments},
  doi       = {10.1103/PhysRevC.89.031601},
  pages     = {031601(R)},
  volume    = {89},
  journal   = {Phys. Rev. C},
  publisher = {American Physical Society},
  year      = {2014},
}

@Article{yu2017,
  author  = {C. Yu and L. Guo},
  title   = {Angular momentum dependence of quasifission dynamics in the reaction $^{48}\mathrm{Ca}+{}^{244}\mathrm{Pu}$},
  doi     = {10.1007/s11433-017-9063-3},
  pages   = {092011},
  volume  = {60},
  journal = {Sci. China-Phys. Mech. Astron.},
  year    = {2017},
}

@Article{sekizawa2019b,
  author    = {Sekizawa, K. and Hagino, K.},
  title     = {Time-dependent {H}artree-{F}ock plus {L}angevin approach for hot fusion reactions to synthesize the ${Z}=120$ superheavy element},
  doi       = {10.1103/PhysRevC.99.051602},
  pages     = {051602(R)},
  volume    = {99},
  journal   = {Phys. Rev. C},
  publisher = {American Physical Society},
  year      = {2019},
}

@Article{pigg2014,
  author  = {D. A. Pigg and A. S. Umar and V. E. Oberacker},
  title   = {{E}ulerian rotations of deformed nuclei for {TDDFT} calculations},
  doi     = {10.1016/j.cpc.2014.02.004},
  pages   = {1410--1414},
  volume  = {185},
  journal = {Comput. Phys. Commun.},
  year    = {2014},
}

@Article{kozulin2010,
  author    = {E. M. Kozulin and G. N. Knyazheva and I. M. Itkis and M. G. Itkis and A. A. Bogachev and L. Krupa and T. A. Loktev and S. V. Smirnov and V. I. Zagrebaev and J. \"Ayst\"o and W. H. Trzaska and V. A. Rubchenya and E. Vardaci and A. M. Stefanini and M. Cinausero and L. Corradi and E. Fioretto and P. Mason and G. F. Prete and R. Silvestri and S. Beghini and G. Montagnoli and F. Scarlassara and F. Hanappe and S. V. Khlebnikov and J. Kliman and A. Brondi and {A. Di} Nitto and R. Moro and N. Gelli and S. Szilner},
  title     = {{I}nvestigation of the reaction $^{64}\mathrm{Ni}+{}^{238}\mathrm{U}$ being an option of synthesizing element 120},
  doi       = {10.1016/j.physletb.2010.02.041},
  pages     = {227--232},
  volume    = {686},
  journal   = {Phys. Lett. B},
  publisher = {Elsevier},
  year      = {2010},
}

@Article{wang2017,
  author    = {Wang, Bing and Wen, Kai and Zhao, Wei-Juan and Zhao, En-Guang and Zhou, Shan-Gui},
  title     = {Systematics of capture and fusion dynamics in heavy-ion collisions},
  doi       = {10.1016/j.adt.2016.06.003},
  pages     = {281--370},
  volume    = {114},
  journal   = {At. Data Nucl. Data Tables},
  publisher = {Elsevier {BV}},
  year      = {2017},
}

@Article{vardaci2019,
  author    = {E. Vardaci and M. G. Itkis and I. M. Itkis and G. Knyazheva and E. M. Kozulin},
  title     = {Fission and quasifission toward the superheavy mass region},
  doi       = {10.1088/1361-6471/ab3118},
  pages     = {103002},
  volume    = {46},
  journal   = {J. Phys. G: Nucl. Part. Phys.},
  publisher = {{IOP} Publishing},
  year      = {2019},
}

@Article{hofmann2016,
  author    = {S. Hofmann and S. Heinz and R. Mann and J. Maurer and G. M\"unzenberg and S. Antalic and W. Barth and H. G. Burkhard and L. Dahl and K. Eberhardt and R. Grzywacz and J. H. Hamilton and R. A. Henderson and J. M. Kenneally and B. Kindler and I. Kojouharov and R. Lang and B. Lommel and K. Miernik and D. Miller and K. J. Moody and K. Morita and K. Nishio and A. G. Popeko and J. B. Roberto and J. Runke and K. P. Rykaczewski and S. Saro and C. Scheidenberger and H. J. Sch\"ott and D. A. Shaughnessy and M. A. Stoyer and P. {Th\"orle-Pospiech} and K. Tinschert and N. Trautmann and J. Uusitalo and A. V. Yeremin},
  title     = {Review of even element super-heavy nuclei and search for element 120},
  doi       = {10.1140/epja/i2016-16180-4},
  pages     = {180},
  volume    = {52},
  journal   = {Eur. Phys. J. A},
  publisher = {Springer Nature},
  year      = {2016},
}

@Article{khuyagbaatar2020,
  author    = {Khuyagbaatar, J. and Yakushev, A. and D\"ullmann, Ch. E. and Ackermann, D. and Andersson, L.-L. and Asai, M. and Block, M. and Boll, R. A. and Brand, H. and Cox, D. M. and Dasgupta, M. and Derkx, X. and Di Nitto, A. and Eberhardt, K. and Even, J. and Evers, M. and Fahlander, C. and Forsberg, U. and Gates, J. M. and Gharibyan, N. and Golubev, P. and Gregorich, K. E. and Hamilton, J. H. and Hartmann, W. and Herzberg, R.-D. and He\ss{}berger, F. P. and Hinde, D. J. and Hoffmann, J. and Hollinger, R. and H\"ubner, A. and J\"ager, E. and Kindler, B. and Kratz, J. V. and Krier, J. and Kurz, N. and Laatiaoui, M. and Lahiri, S. and Lang, R. and Lommel, B. and Maiti, M. and Miernik, K. and Minami, S. and Mistry, A. K. and Mokry, C. and Nitsche, H. and Omtvedt, J. P. and Pang, G. K. and Papadakis, P. and Renisch, D. and Roberto, J. B. and Rudolph, D. and Runke, J. and Rykaczewski, K. P. and Sarmiento, L. G. and Sch\"adel, M. and Schausten, B. and Semchenkov, A. and Shaughnessy, D. A. and Steinegger, P. and Steiner, J. and Tereshatov, E. E. and Th\"orle-Pospiech, P. and Tinschert, K. and Torres De Heidenreich, T. and Trautmann, N. and T\"urler, A. and Uusitalo, J. and Wegrzecki, M. and Wiehl, N. and Van Cleve, S. M. and Yakusheva, V.},
  title     = {Search for elements 119 and 120},
  doi       = {10.1103/PhysRevC.102.064602},
  pages     = {064602},
  volume    = {102},
  journal   = {Phys. Rev. C},
  publisher = {American Physical Society},
  year      = {2020},
}

@Article{sekizawa2020,
  author    = {Sekizawa, Kazuyuki and Ayik, Sakir},
  title     = {Quantal diffusion approach for multinucleon transfer processes in the $^{58,64}\mathrm{Ni}+^{208}\mathrm{Pb}$ reactions: {T}oward the production of unknown neutron-rich nuclei},
  doi       = {10.1103/PhysRevC.102.014620},
  pages     = {014620},
  volume    = {102},
  journal   = {Phys. Rev. C},
  publisher = {American Physical Society},
  year      = {2020},
}

@Article{godbey2019c,
  author    = {Godbey, K. and Guo, Lu and Umar, A. S.},
  title     = {Influence of the tensor interaction on heavy-ion fusion cross sections},
  doi       = {10.1103/PhysRevC.100.054612},
  pages     = {054612},
  volume    = {100},
  journal   = {Phys. Rev. C},
  publisher = {American Physical Society},
  year      = {2019},
}

@Article{chabanat1998a,
  author  = {E. Chabanat and P. Bonche and P. Haensel and J. Meyer and R. Schaeffer},
  title   = {{A S}kyrme parametrization from subnuclear to neutron star densities {P}art {II}. {N}uclei far from stabilities},
  doi     = {10.1016/S0375-9474(98)00180-8},
  pages   = {231--256},
  volume  = {635},
  journal = {Nucl. Phys. A},
  year    = {1998},
}

@Article{hamilton2013,
  author  = {Hamilton, J. H. and Hofmann, S. and Oganessian, {\relax Yu. Ts}.},
  title   = {{S}earch for {S}uperheavy {N}uclei},
  doi     = {10.1146/annurev-nucl-102912-144535},
  pages   = {383--405},
  volume  = {63},
  journal = {Annu. Rev. Nucl. Part. Sci.},
  year    = {2013},
}

@Article{sun2023b,
  author    = {Sun, Xiang-Xiang and Guo, Lu},
  title     = {Microscopic study of the hot-fusion reaction $^{48}\mathrm{Ca}+^{238}\mathrm{U}$ with the constraints from time-dependent {H}artree-{F}ock theory},
  doi       = {10.1103/PhysRevC.107.064609},
  pages     = {064609},
  volume    = {107},
  journal   = {Phys. Rev. C},
  publisher = {American Physical Society},
  year      = {2023},
}

@Article{huang2024b,
  author    = {Huang, Yun and Sun, Xiang-Xiang and Guo, Lu},
  title     = {Role of the tensor force in induced fission of $^{240}\mathrm{Pu}$},
  doi       = {10.1103/PhysRevC.110.064318},
  pages     = {064318},
  volume    = {110},
  journal   = {Phys. Rev. C},
  publisher = {American Physical Society},
  year      = {2024},
}

@Article{li2024c,
  author    = {Li, Liang and Guo, Lu and Godbey, K. and Umar, A. S.},
  title     = {Impact of tensor forces on quasifission product yield distributions},
  doi       = {10.1103/PhysRevC.110.064607},
  pages     = {064607},
  volume    = {110},
  journal   = {Phys. Rev. C},
  publisher = {American Physical Society},
  year      = {2024},
}

@Article{goddard2015,
  author    = {Goddard, P. M. and Stevenson, P. D. and Rios, A.},
  title     = {{F}ission dynamics within time-dependent {H}artree-{F}ock: deformation-induced fission},
  doi       = {10.1103/PhysRevC.92.054610},
  pages     = {054610},
  volume    = {92},
  journal   = {Phys. Rev. C},
  publisher = {American Physical Society},
  year      = {2015},
}

@Article{wang2006,
  author    = {Wang, Ning and Wu, Xizhen and Li, Zhuxia and Liu, Min and Scheid, Werner},
  title     = {Applications of {S}kyrme energy-density functional to fusion reactions for synthesis of superheavy nuclei},
  doi       = {10.1103/PhysRevC.74.044604},
  pages     = {044604},
  volume    = {74},
  journal   = {Phys. Rev. C},
  publisher = {American Physical Society},
  year      = {2006},
}

@Article{hofmann2000,
  author  = {S. Hofmann and G. M\"unzenberg},
  title   = {{T}he discovery of the heaviest elements},
  doi     = {10.1103/RevModPhys.72.733},
  pages   = {733--767},
  volume  = {72},
  journal = {Rev. Mod. Phys.},
  year    = {2000},
}

@Article{stevenson2019,
  author  = {P. D. Stevenson and M. C. Barton},
  title   = {Low--energy heavy-ion reactions and the {S}kyrme effective interaction},
  doi     = {10.1016/j.ppnp.2018.09.002},
  pages   = {142--164},
  volume  = {104},
  journal = {Prog. Part. Nucl. Phys.},
  year    = {2019},
}

@Article{sun2022c,
  author    = {Sun, Xiang-Xiang and Guo, Lu},
  title     = {Effects of the tensor force on low-energy heavy-ion fusion reactions: {A} mini review},
  doi       = {10.1088/1572-9494/ac7e28},
  pages     = {097302},
  volume    = {74},
  journal   = {Commun. Theor. Phys.},
  publisher = {IOP Publishing},
  year      = {2022},
}

@Article{dai2014,
  author    = {Dai, Gao--Feng and Guo, Lu and Zhao, En-Guang and Zhou, Shan--Gui},
  title     = {Dissipation dynamics and spin--orbit force in time--dependent {H}artree--{F}ock theory},
  doi       = {10.1103/PhysRevC.90.044609},
  pages     = {044609},
  volume    = {90},
  journal   = {Phys. Rev. C},
  publisher = {American Physical Society},
  year      = {2014},
}

@Article{oganessian2010,
  author  = {{Yu. Ts. Oganessian} and {F. Sh. Abdullin} and P. D. Bailey and D. E. Benker and M. E. Bennett and S. N. Dmitriev and J. G. Ezold and J. H. Hamilton and R. A. Henderson and M. G. Itkis and {Yu. V. Lobanov} and A. N. Mezentsev and K. J. Moody and S. L. Nelson and A. N. Polyakov and C. E. Porter and A. V. Ramayya and F. D. Riley and J. B. Roberto and M. A. Ryabinin and K. P. Rykaczewski and R. N. Sagaidak and D. A. Shaughnessy and I. V. Shirokovsky and M. A. Stoyer and V. G. Subbotin and R. Sudowe and A. M. Sukhov and {Yu. S. Tsyganov} and V. K. Utyonkov and A. A. Voinov and G. K. Vostokin and P. A. Wilk},
  title   = {{S}ynthesis of a {N}ew {E}lement with {A}tomic {N}umber ${Z}=117$},
  doi     = {10.1103/PhysRevLett.104.142502},
  pages   = {142502},
  volume  = {104},
  journal = {Phys. Rev. Lett.},
  year    = {2010},
}

@Article{zagrebaev2015,
  author    = {V.I. Zagrebaev and W. Greiner},
  title     = {Cross sections for the production of superheavy nuclei},
  doi       = {10.1016/j.nuclphysa.2015.02.010},
  pages     = {257--307},
  volume    = {944},
  journal   = {Nucl. Phys. A},
  publisher = {Elsevier {BV}},
  year      = {2015},
}

@Article{wu2022,
  author  = {Zhenji Wu and Lu Guo and Zhong Liu and Guangxiong Peng},
  title   = {Production of proton-rich nuclei in the vicinity of $^{100}\mathrm{Sn}$ via multinucleon transfer reactions},
  doi     = {10.1016/j.physletb.2022.136886},
  pages   = {136886},
  volume  = {825},
  journal = {Phys. Lett. B},
  year    = {2022},
}

@Article{huang2024,
  author    = {Huang, Yun and Sun, Xiang-Xiang and Guo, Lu},
  title     = {Fission fragment distributions within time-dependent density functional theory},
  doi       = {10.1140/epja/s10050-024-01326-2},
  pages     = {100},
  volume    = {60},
  journal   = {Eur. Phys. J. A},
  publisher = {Springer Science and Business Media LLC},
  year      = {2024},
}

@Article{li2019,
  author  = {Li, XiaoYu and Wu, ZhenJi and Guo, Lu},
  title   = {Entrance-channel dynamics in the reaction $^{40}\mathrm{Ca}+{}^{208}\mathrm{Pb}$},
  doi     = {10.1007/s11433-019-9435-x},
  pages   = {122011},
  volume  = {62},
  journal = {Sci. China-Phys. Mech. Astron.},
  year    = {2019},
}

@Article{morita2004,
  author    = {Kosuke Morita and Kouji Morimoto and Daiya Kaji and Takahiro Akiyama and Sin-ichi Goto and Hiromitsu Haba and Eiji Ideguchi and Rituparna Kanungo and Kenji Katori and Hiroyuki Koura and Hisaaki Kudo and Tetsuya Ohnishi and Akira Ozawa and Toshimi Suda and Keisuke Sueki and HuShan Xu and Takayuki Yamaguchi and Akira Yoneda and Atsushi Yoshida and YuLiang Zhao},
  title     = {Experiment on the {S}ynthesis of {E}lement 113 in the {R}eaction $^{209}\mathrm{Bi}(^{70}\mathrm{Zn},n)^{278}113$},
  doi       = {10.1143/jpsj.73.2593},
  pages     = {2593--2596},
  volume    = {73},
  journal   = {J. Phys. Soc. Japan},
  publisher = {Physical Society of Japan},
  year      = {2004},
}

@Article{back1983,
  author    = {B. B. Back and R. R. Betts and K. Cassidy and B. G. Glagola and J. E. Gindler and L. E. Glendenin and B. D. Wilkins},
  title     = {{E}xperimental {S}ignatures of {Q}uasifission {R}eactions},
  doi       = {10.1103/PhysRevLett.50.818},
  pages     = {818--821},
  volume    = {50},
  journal   = {Phys. Rev. Lett.},
  publisher = {American Physical Society},
  year      = {1983},
}

@Article{li2022,
  author  = {Liang Li and Lu Guo and K. Godbey and A. S. Umar},
  title   = {Impact of tensor force on quantum shell effects in quasifission reactions},
  doi     = {10.1016/j.physletb.2022.137349},
  pages   = {137349},
  volume  = {833},
  journal = {Phys. Lett. B},
  year    = {2022},
}

@Article{oganessian2009,
  author    = {Oganessian, Yu. Ts. and Utyonkov, V. K. and Lobanov, Yu. V. and Abdullin, F. Sh. and Polyakov, A. N. and Sagaidak, R. N. and Shirokovsky, I. V. and Tsyganov, Yu. S. and Voinov, A. A. and Mezentsev, A. N. and Subbotin, V. G. and Sukhov, A. M. and Subotic, K. and Zagrebaev, V. I. and Dmitriev, S. N. and Henderson, R. A. and Moody, K. J. and Kenneally, J. M. and Landrum, J. H. and Shaughnessy, D. A. and Stoyer, M. A. and Stoyer, N. J. and Wilk, P. A.},
  title     = {Attempt to produce element 120 in the $^{244}\mathrm{Pu}+^{58}\mathrm{Fe}$ reaction},
  doi       = {10.1103/PhysRevC.79.024603},
  pages     = {024603},
  volume    = {79},
  journal   = {Phys. Rev. C},
  publisher = {American Physical Society},
  year      = {2009},
}

@Article{lee2024b,
  author    = {Lee, H. and McGlynn, P. and Simenel, C.},
  title     = {Shell effects in quasifission in reactions forming the $^{226}\mathrm{Th}$ compound nucleus},
  doi       = {10.1103/PhysRevC.110.024606},
  pages     = {024606},
  volume    = {110},
  journal   = {Phys. Rev. C},
  publisher = {American Physical Society},
  year      = {2024},
}

@Article{colo2007,
  author  = {G. Col\`{o} and H. Sagawa and S. Fracasso and P. F. Bortignon},
  title   = {Spin--orbit splitting and the tensor component of the {S}kyrme interaction},
  doi     = {10.1016/j.physletb.2007.01.033},
  pages   = {227--231},
  volume  = {646},
  journal = {Phys. Lett. B},
  year    = {2007},
}

@Article{oganessian2007,
  author  = {{Yuri Oganessian}},
  title   = {{H}eaviest nuclei from $^{48}${C}a-induced reactions},
  doi     = {10.1088/0954-3899/34/4/R01},
  pages   = {R165--R242},
  volume  = {34},
  journal = {J. Phys. G: Nucl. Part. Phys.},
  year    = {2007},
}

@Article{oganessian2006,
  author    = {Oganessian, {\relax Yu. Ts}. and Utyonkov, V. K. and Lobanov, {\relax Yu. V}. and Abdullin, {\relax F. Sh}. and Polyakov, A. N. and Sagaidak, R. N. and Shirokovsky, I. V. and Tsyganov, {\relax Yu. S}. and Voinov, A. A. and Gulbekian, G. G. and Bogomolov, S. L. and Gikal, B. N. and Mezentsev, A. N. and Iliev, S. and Subbotin, V. G. and Sukhov, A. M. and Subotic, K. and Zagrebaev, V. I. and Vostokin, G. K. and Itkis, M. G. and Moody, K. J. and Patin, J. B. and Shaughnessy, D. A. and Stoyer, M. A. and Stoyer, N. J. and Wilk, P. A. and Kenneally, J. M. and Landrum, J. H. and Wild, J. F. and Lougheed, R. W.},
  title     = {Synthesis of the isotopes of elements 118 and 116 in the $^{249}\mathrm{Cf}$ and $^{245}\mathrm{Cm}+{}^{48}\mathrm{Ca}$ fusion reactions},
  doi       = {10.1103/PhysRevC.74.044602},
  pages     = {044602},
  volume    = {74},
  journal   = {Phys. Rev. C},
  publisher = {American Physical Society},
  year      = {2006},
}

@Article{hagino1999,
  author    = {K. Hagino and N. Rowley and A. T. Kruppa},
  title     = {A program for coupled-channel calculations with all order couplings for heavy-ion fusion reactions},
  doi       = {10.1016/s0010-4655(99)00243-x},
  pages     = {143--152},
  volume    = {123},
  journal   = {Comput. Phys. Commun.},
  publisher = {Elsevier},
  year      = {1999},
}

@Article{guo2018,
  author  = {Lu Guo and C\'{e}dric Simenel and Long Shi and Chong Yu},
  title   = {The role of tensor force in heavy-ion fusion dynamics},
  doi     = {10.1016/j.physletb.2018.05.066},
  pages   = {401--405},
  volume  = {782},
  journal = {Phys. Lett. B},
  year    = {2018},
}

@Article{guo2012,
  author  = {Lu Guo and Takashi Nakatsukasa},
  title   = {{T}ime-dependent {H}artree-{F}ock studies of the dynamical fusion threshold},
  doi     = {10.1051/epjconf/20123809003},
  pages   = {09003},
  volume  = {38},
  journal = {{EPJ} {W}eb {C}onf.},
  year    = {2012},
}

@Article{washiyama2015,
  author    = {Washiyama, Kouhei},
  title     = {{M}icroscopic analysis of fusion hindrance in heavy nuclear systems},
  doi       = {10.1103/PhysRevC.91.064607},
  pages     = {064607},
  volume    = {91},
  journal   = {Phys. Rev. C},
  publisher = {American Physical Society},
  year      = {2015},
}

@Article{wu2019,
  author    = {Wu, Zhenji and Guo, Lu},
  title     = {Microscopic studies of production cross sections in multinucleon transfer reaction $^{58}\mathrm{Ni}+^{124}\mathrm{Sn}$},
  doi       = {10.1103/PhysRevC.100.014612},
  pages     = {014612},
  volume    = {100},
  journal   = {Phys. Rev. C},
  publisher = {American Physical Society},
  year      = {2019},
}

@Article{swiatecki2005,
  author    = {W.J.\'{S}wi\c{a}tecki and K.Siwek--Wilczy\'nska and J.Wilczy\'nski},
  title     = {{F}usion by diffusion. {II}. {S}ynthesis of transfermium elements in cold fusion reactions},
  doi       = {10.1103/PhysRevC.71.014602},
  pages     = {014602},
  volume    = {71},
  journal   = {Phys. Rev. C},
  publisher = {American Physical Society},
  year      = {2005},
}

@Article{jachimowicz2021,
  author    = {Jachimowicz, P. and Kowal, M. and Skalski, J.},
  title     = {Properties of heaviest nuclei with 98$\leq$ {Z} $\leq $126 and 134 $\leq$ {N} $\leq $192},
  doi       = {10.1016/j.adt.2020.101393},
  issn      = {0092-640X},
  pages     = {101393},
  volume    = {138},
  journal   = {Atomic Data and Nuclear Data Tables},
  month     = mar,
  publisher = {Elsevier BV},
  year      = {2021},
}

@Article{wakhle2014,
  author    = {Wakhle, A. and Simenel, C. and Hinde, D. J. and Dasgupta, M. and Evers, M. and Luong, D. H. and du Rietz, R. and Williams, E.},
  title     = {{I}nterplay between {Q}uantum {S}hells and {O}rientation in {Q}uasifission},
  doi       = {10.1103/PhysRevLett.113.182502},
  pages     = {182502},
  volume    = {113},
  journal   = {Phys. Rev. Lett.},
  publisher = {American Physical Society},
  year      = {2014},
}

@Article{simenel2018,
  author    = {C. Simenel and A. S. Umar},
  title     = {Heavy-ion collisions and fission dynamics with the time--dependent {H}artree-{F}ock theory and its extensions},
  doi       = {10.1016/j.ppnp.2018.07.002},
  pages     = {19--66},
  volume    = {103},
  journal   = {Prog. Part. Nucl. Phys.},
  publisher = {Elsevier},
  year      = {2018},
}

@Article{godbey2019,
  author    = {Godbey, K. and Umar, A. S. and Simenel, C},
  title     = {Deformed shell effects in ${}^{48}\mathrm{Ca}+{}^{249}\mathrm{Bk}$ quasifission fragments},
  doi       = {10.1103/PhysRevC.100.024610},
  pages     = {024610},
  volume    = {100},
  journal   = {Phys. Rev. C},
  publisher = {American Physical Society},
  year      = {2019},
}
%
%
%
%

\end{document}